\documentclass{nature}

\usepackage{braket}
\usepackage{graphicx}
\usepackage{amsmath}

\usepackage{color}

\title{Colossal Bulk Photovoltaic Effect in a Weyl Semimetal} 

\author{Gavin B. Osterhoudt,$^{1}$ Laura K. Diebel,$^{1}$ Mason J. Gray,$^{1}$ Xu Yang,$^{1}$ John Stanco,$^{1}$\\Xiangwei Huang,$^{2}$ Bing Shen,$^{3}$ Ni Ni,$^{3}$ Philip J.W. Moll,$^{2,4}$ Ying Ran,$^{1}$ Kenneth S. Burch$^{1*}$}

\begin{document}

\maketitle

\begin{affiliations}
 \item Department of Physics, Boston College, 140 Commonwealth Avenue, Chestnut Hill, MA 02467, USA
 \item Max-Planck-Institute for Chemical Physics of Solids, 01187 Dresden, Germany
 \item Department of Physics and Astronomy and California NanoSystems Institute, University of California Los Angeles, CA
 90095, USA
 \item EPFL, IMX, CH-1015 Lausanne, Switzerland
\end{affiliations}


\begin{abstract}
Broadband, efficient and fast conversion of light to electricity is crucial for sensing and clean energy. Here we reveal the largest observed bulk photo-voltaic effect (BPVE), an intrinsic mechanism predicted to be ultrafast and exceed the Shockley-Quiesser limit. This discovery results from combining recent developments in the connection of BPVE to topology, Weyl semimetals and focused-ion beam fabrication. Our room temperature observation of the first BPVE in the mid-IR, is enabled by microscopic devices of the Weyl semimetal TaAs. Detailed symmetry analysis enables unambiguous separation of this response from competing photo-thermal effects. The size and wavelength range of the shift current offers new opportunities in optical detectors, clean energy, and topology, while directly demonstrating the utility of Weyl semimetals for applications.
\end{abstract}

Converting light to electricity is crucial for clean energy, imaging, communications, chemical and biological sensing\cite{brongersma2015,krishna2005JPD}. Recently, interest has been renewed in the non-linear generation of light due to its connection to electron topology and potential for overcoming extrinsic limitations\cite{cook2017NComm,yang2018Sci,tan2016CompMat,morimoto2016SciAdv,ganichev2014PhRep,okada2016PRB,pan2017NatComm,Dhara:2015di}. For example, the use of built-in electric fields present in $p$-$n$ junctions is limited by the principle of detailed balance vis-\'{a}-vis the Shockley-Quiesser limit\cite{shockley1961}. Alternatively, optically induced thermal gradients producing currents via the Seebeck effect require careful balance of the optical, electronic, and thermal material properties. An important alternative is the bulk photovoltaic effect (BPVE), a non-linear mechanism where direct acceleration of the quasi-particles suggests an ultrafast response with fewer limitations on efficiency or maximum open circuit voltage\cite{tan2016CompMat,gan2008ultraviolet,cook2017NComm}. However, the non-linear nature of the BPVE has generally limited its size to be far below a technologically relevant regime. Furthermore its observation primarily in ferroelectric insulators and semiconductors has relegated the effect to a narrow range of wavelengths\cite{tan2016CompMat,zenkevich2014PRB,yang2018Sci}.

The focus of BPVE studies on ferrorelectrics resulted from its original observation there and its attribution to built-in fields\cite{chynoweth1956}. However, today BPVEs are understood through the lens of non-linear optics\cite{sipe2000PRB,ganichev2014PhRep} and topology, namely the evolution of the wavefunction in momentum space\cite{cook2017NComm,morimoto2016SciAdv,ying2017,chan2017PRB,Golub2017}. Of particular interest are shift currents that result from a change in the center of the Wannier function of the particle upon interband optical excitation. These are intimately linked to a measure of the topology of the band structure, the expectation value of the position operator in the crystal unit cell (Berry connection). Thus upon excitation between bands the change in Berry connection and phase of the velocity operator produces a real space shift in the particle position, generating a time dependent dipole moment and ultimately a dc current. This need to generate a time dependent dipole results in shift currents only emerging from linearly polarized light. Alternatively, circularly polarized light can produce a time dependent current, which via scattering becomes a dc response known as the injection current\cite{sipe2000PRB,chan2017PRB}. Indeed as originally proposed in reference 17, in a WSM these injection currents are intimately tied to the electron chirality. In our experiments performed at room temperature, we generally do not observe a circularly polarized response, consistent with short scattering times. From a classical electrodynamics point of view, this direct generation of a dc photocurrent by finite frequency electric fields is only possible via a non-linear optical conductivity, $\sigma_{ijk}^{(2)}$\cite{sipe2000PRB,ganichev2014PhRep,ma2017Nat,okada2016PRB,chan2017PRB}. By symmetry this term is only possible in the absence of a center of inversion, and for certain current directions contains products of electric fields along orthogonal crystal axes. As depicted in Fig.~1A, this allows shift currents to be zero for particular polarizations of light. This is quite distinct from photothermal effects that are maximized for these polarizations as they only depend on the absorbed power\cite{xu2010NLett,ma2017Nat,okada2016PRB,pan2017NatComm}.

The requirement for broken inversion symmetry has been put to use in inducing a BPVE in Si and SrTiO$_{3}$\cite{yang2018Sci}, but can also lead to Weyl semimetals (WSM) with a singular Berry connection. Indeed, the breaking of inversion combined with spin-orbit coupling generates a three dimensional linear dispersion with quasi-particles of well-defined chirality\cite{armitage2017RMP}. Recently a number of compounds have been confirmed to be in this class of materials, with experimental and theoretical evidence for the Weyl nodes containing a singular Berry connection\cite{hding2015NPhys,yang2015NPhys,hasan2015Sci,shuang2016NComm}. The novel Berry connection of these materials has led to suggestions of exploiting WSM to generate novel optical effects\cite{ying2017,zhang2018arXiv,chan2017PRB,Golub2017}, as well as the experimental demonstration of non-linear optics to measure their chirality at low temperatures\cite{ma2017Nat,chan2017PRB}. Furthermore, the gapless electronic spectrum of a WSM  opens the possibility of extending BPVE into previously inaccessible energy ranges\cite{ying2017,chan2017PRB}. In this paper we utilize the unique topology of the WSM TaAs and new fabrication techniques to reveal a room temperature, colossal BPVE in an entirely new wavelength regime.

\section*{Optimization of Weyl semimetal photocurrent measurements}

TaAs is a member of the (Ta,Nb)(P,As) family of compounds whose topological classification as a WSM has been extensively confirmed. The separation between the nodes of different chiralities is largest in TaAs, and thus BPVE is expected to occur throughout a wide range in energy. As shown in the supplemental, and consistent with previous studies\cite{ma2017Nat,hding2015NPhys,yang2015NPhys,hasan2015Sci,shuang2016NComm}, our devices have a Fermi level close to the Weyl nodes. These materials generally possess two inequivalent Weyl nodes, through which mirror symmetries in the $ab$-plane produce a total of 24 nodes throughout the Brillouin zone\cite{weng2015PRX,armitage2017RMP}. Previous studies of the nonlinear optical properties of TaAs have revealed the largest second harmonic generation mechanism\cite{wu2017NPhys}, as well as a circular photogalvanic effect likely related to the electron chirality\cite{ma2017Nat,chan2017PRB,Golub2017}. The latter study was performed on a bulk crystal limiting the temperature range at which non-linear effects were observable due to the dominance of thermal effects and resistive losses at ambient temperatures. This was explained as the natural result of two competing length scales: the first being the the spot size versus the contact separation; and the second the ratio of the penetration depth to the device thickness. These result from the nonlinear photocurrent generation occurring only in the illuminated region of the device, with the resulting current spreading out until it reaches the contacts (see Fig.~1B). If the beam size is much smaller than the contact separation, the thickness much greater than the penetration depth, or both, the measured photocurrent will be strongly suppressed by resistive loses. To mitigate this we produced a microscopic device by utilizing recent developments in focused-ion-beam (FIB) based fabrication\cite{moll2018AR}. As shown by a false-color SEM image of one such device in Fig.~1C, these have a thickness roughly 2-3 times the penetration depth ($\sim$250 nm)\cite{xu2016PRB}. Furthermore, the lateral dimensions of the device are approximately 11$\times$2 $\mu$m, which are smaller than the minimum spot size achievable using our focusing optics (approximately 30 $\mu$m in diameter). As discussed in the supplemental materials, this reduces the effect of thermal gradients but does not completely eliminate them.

\begin{figure}
    \centering
    \includegraphics[width=0.96\textwidth]{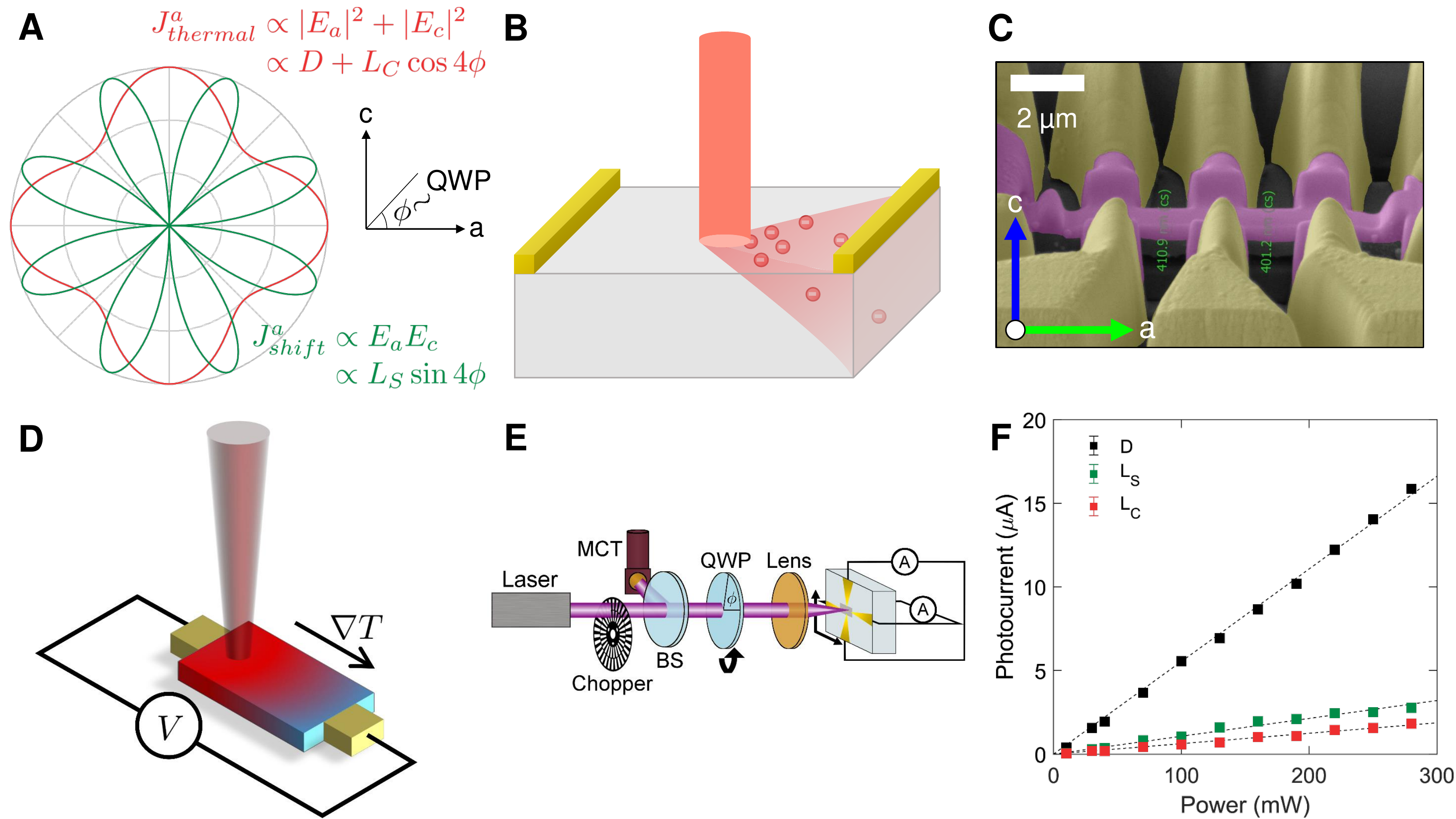}
    \caption{\textbf{Symmetry and Photocurrent Generation} \textbf{(A)} Polarization dependence of thermal and shift current photocurrent contributions. The radius and angle of the polar plot correspond to the magnitude of the photocurrents and the angle the fast-axis of the QWP makes with the $x$-axis respectively. The intrinsic and extrinsic responses have a different phase as can be seen by the out-of-phase minima and maxima. \textbf{(B)} Resistive losses occur due to the large sample size compared to the small generation region. \textbf{(C)} False color SEM image of a microcopic TaAs (purple) device with Au (yellow) contacts. The device is nearly three times thicker than the penetration depth of the light used ($\approx 750$ nm), with an area 20 times smaller than the spot. This simultaneously minimizes thermal response and resistive losses. \textbf{(D)} Diagram of the photo-thermoelectric effect. A light induced temperature gradient gives rise to a potential difference across the device. \textbf{(E)} Schematic of the experimental setup with control of the polarization state via rotation of the quarter wave plate (QWP). \textbf{(F)} Linear power dependence of QWP angle independent ($D$), four-fold intrinsic ($L_{S}$) and thermal ($L_{C}$) terms.}
    \label{fig:Fig1}
\end{figure}

Despite designing the device to minimize their influence, simulations (see supplemental) indicate extraneous currents arising from thermal effects are still expected. The primary mechanism responsible for this is the photo-thermoelectric effect (PTE), illustrated in Fig.~1D, whereby laser induced thermal gradients produce currents through the Seebeck effect\cite{xu2010NLett,ma2017Nat,okada2016PRB}. Beyond the device design, one may hope to  eliminate the PTE by sample positioning alone. However, we found they are typically of the same magnitude as the shift current response, and thus it is difficult to optimize the position based on photocurrent signal alone. Fortunately a detailed analysis of the symmetry properties of these effects (see supplemental) allows for a separation of the PTE response from the shift current\cite{okada2016PRB,pan2017NatComm}. Specifically, the PTE depends solely on the intensity and thus its polarization dependence arises from differences in the absorption of light along various crystal axes. This results in the PTE photocurrent being proportional to $ \alpha_{i} |E_{i}^{2}| $, where $ \alpha_{i}$ is the absorption along a specific axis and the summation over $i$ is implied. On the contrary, the shift current requires an even number of axes to be non-inversion breaking. For the crystal $a$-axis with mirror symmetry, this leads to a dependence proportional to $E_{a} E_{c}^{*} + E_{c} E_{a}^{*}$, while along the $c$-axis the dependence is proportional to $|E_{i}|^{2}$. This implies for photocurrents measured along the $a$-axis, the PTE and shift currents will have exactly opposite polarization dependence. Namely, the PTE is maximized for polarizations purely parallel to the crystal axes, while the shift currents are maximized when the electric field contains components along both axes (see Fig.~1a). 

In order to excite, measure, and separate different symmetry components of the photocurrent response we employed a custom built opto-electronic setup (see Fig.~1E). A CO$_{2}$ laser operating at 10.6\ $\mu$m (117 meV) was selected to minimize reflective losses\cite{xu2016PRB,kimura2017PRB} while still probing electronic excitations including the  Weyl cone\cite{xu2016PRB,ma2017Nat,kimura2017PRB} (see supplemental). The samples were mounted on a motorized $xyz$-translation stage. This provides accurate positioning of the sample and the spatial variation of the photo-response crucial for measuring the spot size and minimizing the PTE (see supplemental). To access the different required polarizations a quarter-wave plate (QWP) was used, similar to previous studies\cite{ma2017Nat,ganichev2014PhRep,okada2016PRB,dongsun2018}. When the incident light is polarized parallel or perpendicular to the fast-axis of the QWP (0 or 90$^{\circ}$ respectively), the polarization is unaltered. Upon rotating the QWP to $\pm$45$^{\circ}$, a phase delay of $\mp \pi/2$ is produced between the two perpendicular components leading to right or left circularly polarized light. At intermediate angles the light becomes elliptically polarized, containing a superposition of two perpendicularly polarized components. This is crucial since some of the shift currents require the mixing of two perpendicular components of light. Taking the angle that the fast axis of the QWP makes with the horizontal as $\phi$, one can use Jones matrices to show that, as expected, the PTE response consists of a constant term and a polarization dependent term that depends on $\cos(4\phi)$ (see Fig.~1a and supplemental for details). This polarization dependence is expected to be the same for photocurrents along both crystal axes, and is identical in form to the shift current along the $c$-axis. On the other hand, due to the requirement of an electric field component along the inversion breaking axis, the shift current response along the $a$-axis depends on $\sin(4\phi)$. This difference in phase is critical to the identification of shift current contributions, so great care was taken in identifying the orientation of the QWP. The variation with QWP angle of the reflectance measured by the MCT detector allowed us to precisely identify the angle at which the QWP fast axis was aligned with the horizontal (see supplemental).

\begin{figure}
    \centering
    \includegraphics[width=0.96\textwidth]{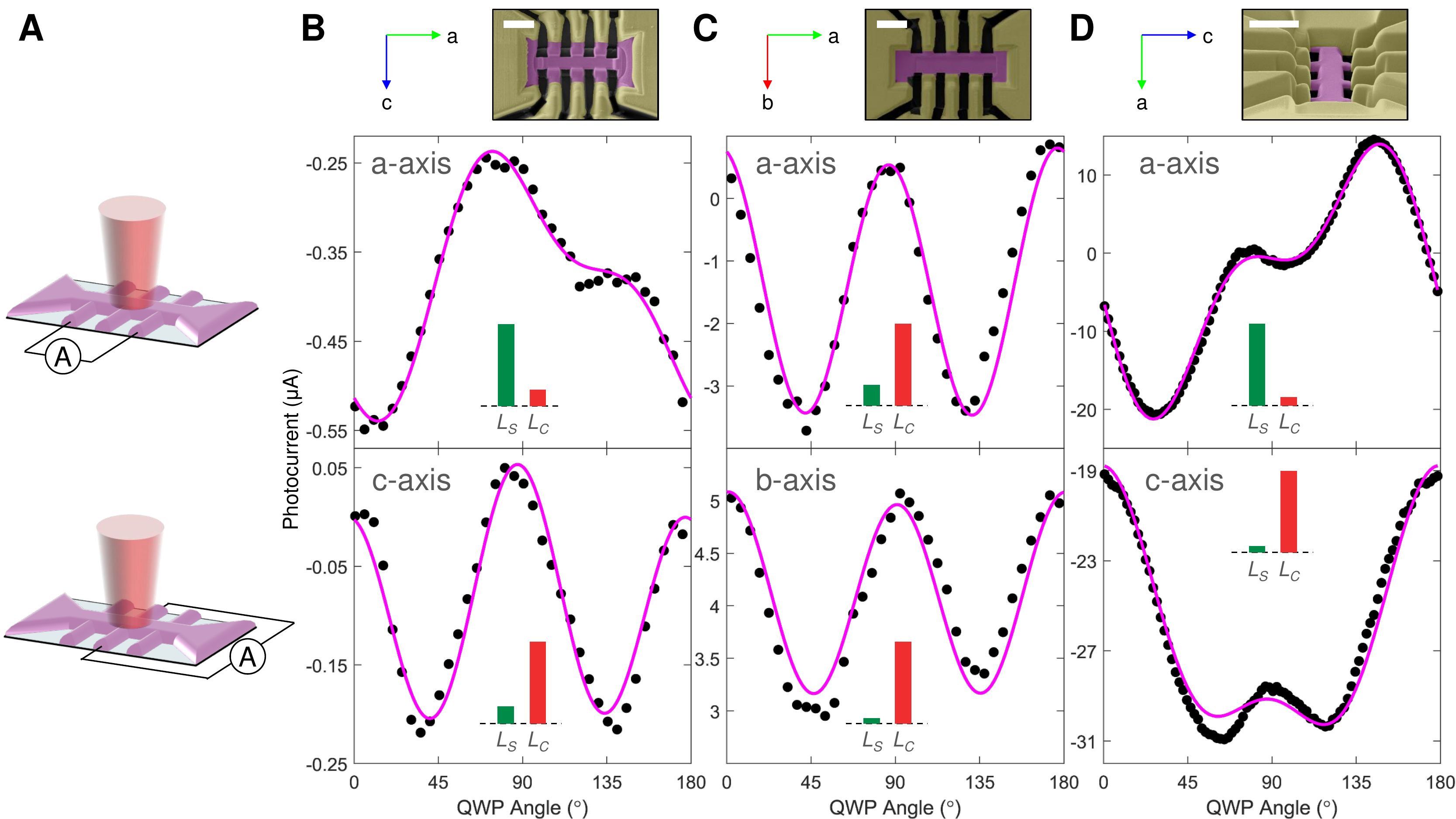}
    \caption{\textbf{Polarization Dependent Photocurrent} \textbf{(A)} Diagram of the device indicating the axis along which the current is probed. \textbf{(B,C,D)} Upper panel, SEM image of the measured device and crystal orientation. Scale bars are 5 $\mu$m. Middle/Lower panel, data (black dots) for photocurrent measured along the longitudinal/transverse axis of the Hall bar. As expected for devices with $c$-axis in the plane, the current along the $a$-axis is dominated by shift current response ($\sin(4\phi)$ - $L_{S}$). In all other cases the term primarily due to thermal response ($\cos(4 \phi)$ - $L_{C}$), is largest. These are determined by fitting (magenta lines) with all possible terms including shift, photo-thermal, and those coming from other effects such as anisotropic absorption in the QWP (for more explanation see the supplemental).}
    \label{fig:Fig2}
\end{figure}

\section*{Separating shift current from the photothermal effects}

We now turn to the results of our photocurrent measurements on three different devices, presented in Fig.~2. An $ac$-face device is shown in panel B, an $ab$-face device in panel C, and a second $ac$-face device rotated 90$^{\circ}$ in panel D. Let us consider the results from the first $ac$-device shown in Fig.~2B. As expected for a dominant shift current response, along the $a$-axis the four-fold sine term ($L_{S}$) was nearly five times larger than the cosine term ($L_{C}$, see inset bars). In comparison, along the broken inversion axis both the shift current and PTE should only be proportional to the sum of the squares of the electric field and we primarily observe a cosine term. The small amount of the sine term in this direction results from an approximately 13$^{\circ}$ difference between the $a$-axis of the device and the horizontal. Details of this estimation are provided in the supplemental.

To confirm the presence of a large shift current along the $a$-axis as indicated by the four-fold sine term, we measured a nearly identical device with an optical $ab$-face. Due to the lack of an in-plane inversion breaking axis this device should only exhibit PTE, as a second-order photocurrent response is excluded by symmetry. As shown in Fig.~2C the four-fold $a$-axis response reveals a sine term which is less than one third the cosine term. In fact, the response along each axis of this $ab$-face device is markedly similar, which substantiates our previous assertion that only thermal terms should be present in such a device.

As a check of the polarization dependence, we rotated a third TaAs device (with an $ac$-face) to be oriented 90$^{\circ}$ from the previous measurement configurations. Such a rotation should change the sign of the shift current term when compared to the original orientation, but not the relative phase (i.e. cosine versus sine). In Fig.~2D we show the photocurrent measured in the rotated state, which like the other $ac$-device reveals a response along the $a$-axis dominated by the sine term. Comparing the sign of the L$_{S}$ term to the one measured when the device was not rotated (see Supplemental Table~1) we see that the sign does indeed change, in agreement with what is expected by symmetry.

Both the PTE and second-order non-linear effects are expected to have a linear dependence on the input power since both of them rely on the square of the electric field. The measured power dependence of the polarization independent, sine, and cosine terms from an $ac$-face device are shown in Fig.~1F. Over the range of powers measured each term is approximately linear, incompatible with a higher order nonlinear mechanism seen in another Weyl system\cite{dongsun2018} and consistent with these generation mechanisms (see supplemental). Other possible built-in electric field sources that would contribute to a third-order response, yet leave the power dependence linear, are eliminated by symmetry requirements on the forms such responses could take.

\section*{Origin and relative size of the BPVE in TaAs}

To further confirm we have extracted the intrinsic shift current response, we compare our experimental results with band structure calculations, detailed in the supplemental. The parameters of the calculation are in good agreement with previously the established band structure of TaAs\cite{kimura2017PRB} as well as our measurement conditions. Specifically, accounting for the ambient temperature and the electron ($n_{e} = 8.08 \times 10^{17}$) and hole ($n_{h} = 9.35 \times 10^{17}$ cm$^{-3}$) concentrations (as measured via Hall effect), which places the Fermi level close to the Weyl node, we calculated $\sigma_{aac} = 201\ \mu$A/V$^{2}$. At first glance this is significantly larger than our apparent measured value of $\sigma_{aac} = 34 \pm 3.7\ \mu$A/V$^{2}$. However, the experimental value does not account for reflective losses. Given the large reflectance of TaAs at 10.6 $\mu$m we find the actual $\sigma_{aac} = 154 \pm 17\ \mu$A/V$^{2}$. Noting some uncertainty in the calculated theoretical and experimental values we find good agreement between the two. In our theoretical calculations, based on a 32 band tight binding model, the valence and conduction bands constituting the Weyl nodes are labeled as band-16 and band-17 respectively. At the experimental photon-energy, there are two inter-band transitions: from bands 16 to 17, and from 16 to 18 (band-18 is the adjacent, higher energy conduction band). We find the contributions to $\sigma_{aac}$ to be $251 \mu$A/V$^{2}$ and $-50\mu$A/V$^{2}$, respectively. Thus our calculations demonstrate the primary contribution is from the Weyl nodes. 

Having established the intrinsic contribution to the shift current, we now turn to compare it with previous measurements. Since shift currents occur for interband transitions, it is important to account for the absorption in the material that reduces the effective depth. This is typically accomplished by calculating the Glass coefficient (measured responsivity divided by the absorption coefficient)\cite{glass1974APL,tan2016CompMat}. It thus accounts for the measured current primarily being generated within a penetration depth. Using our measured values and the absorption coefficient at 10.6\ $\mu$m we calculate $G=1.65 \pm 0.18 \times 10^{-7}$ cm/V. This value is plotted alongside other reported values for $G$ in Fig.~3. We first note that to the best of our knowledge this is the first report of a BPVE in the mid-IR, thus greatly expanding its potential for utility in thermal energy conversion and sensing. Perhaps more relevant to its potential utility, the $G$ we measured here is nearly an order of magnitude larger than the recently reported giant BPVE in BaTiO$_{3}$\cite{zenkevich2014PRB}. This mid-IR colossal BPVE observed in TaAs is enabled by the gapless nature and anomalous Berry connection present in WSMs. Indeed most BPVE reports are in ferroelectric insulators or strained semiconductors with band gaps in the visible range\cite{tan2016CompMat,zenkevich2014PRB,yang2018Sci}. Noting that our reported $G$ for TaAs does not account for reflective losses, we anticipate much larger non-linear photocurrent generation is possible with anti-reflective coatings. Furthermore, as shown in the supplemental materials our calculations of the shift current response over a range of photon energies ($\sim$25-350 meV) reveal a significant enhancement at longer wavelengths, consistent with the presence of the Weyl nodes at lower energies and previous model calculations\cite{ying2017}. Thus the results we present here open up a previously inaccessible energy regime and provide a likely route to the successful use of BPVE and WSM.

\begin{figure}
    \centering
    \includegraphics[width=\textwidth]{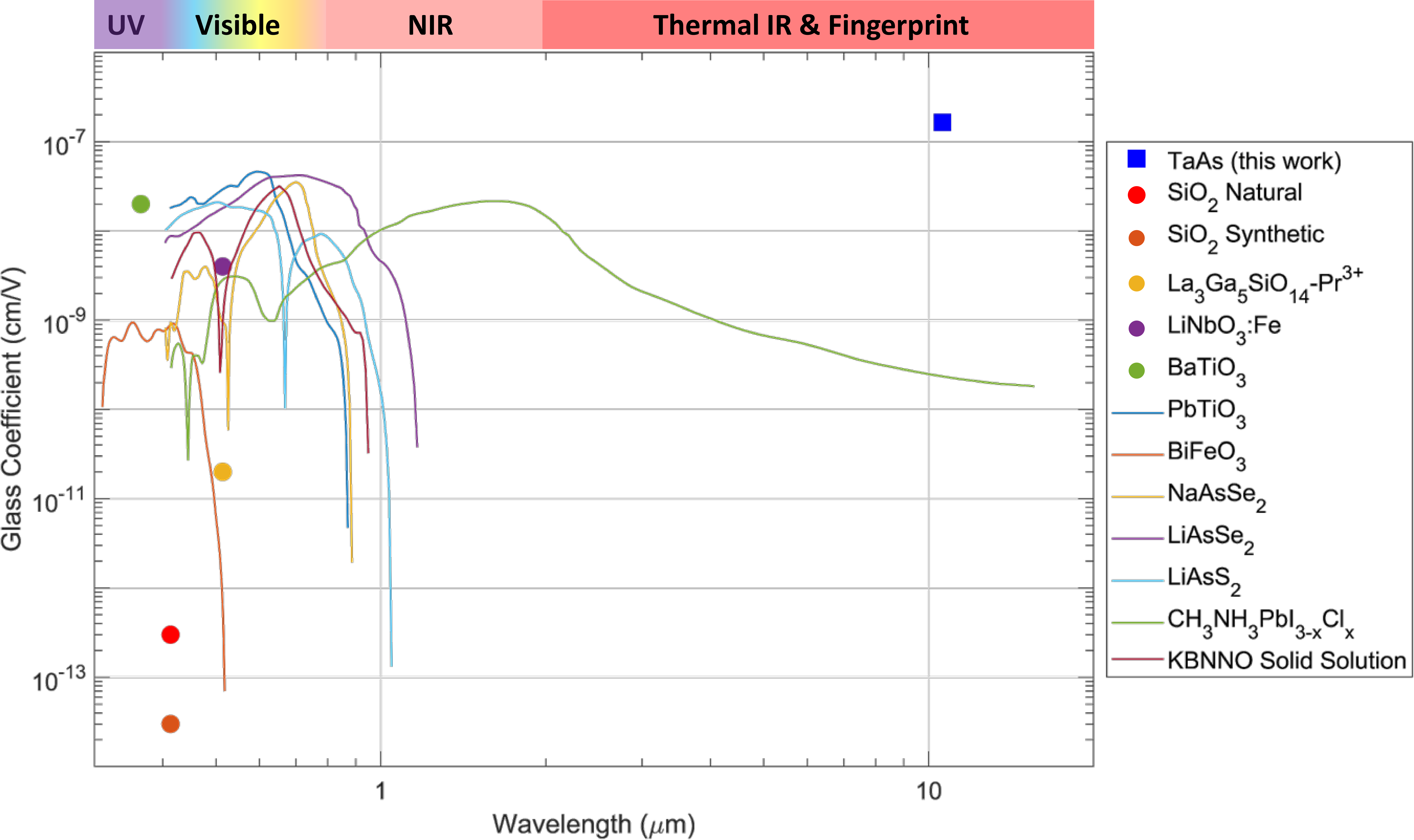}
    \caption{\textbf{Table of Glass Coefficients} The experimentally measured (circles) and calculated (lines) Glass coefficient for various ferroelectric materials. Due to its unique anomalous Berry connection, TaAs (blue square) reveals a nearly one order of magnitude larger response. The gapless nature of this Weyl semimetal enables broadband response in an entirely new regime relevant to thermal, chemical and biological signatures.}
    \label{fig:Fig3}
\end{figure}

\begin{methods}

\noindent
\textbf{TaAs crystal growth and alignment.}
Single crystals of TaAs were grown using the chemical vapor transport method with iodine as the transport agent. The precursor of TaAs powder was first made by solid-state reaction. 3 g of TaAs powder together with iodine pieces were loaded into a 22-cm long quartz tube with an inner diameter of 14 mm and an outer diameter of 18 mm. The density of iodine inside the quartz tube was 18 mg/cm$^3$. The quartz tube was then sealed under vacuum and put into a two-zone furnace with the precursor end at 1050 $^{\circ}$C and the other end at 950 $^{\circ}$C. The whole growth took around 4 weeks.

\noindent
\textbf{Device Fabrication.}
TaAs devices are fabricated by starting from the as-grown single crystals. A Focused Ion Beam (FIB) is used to cut a micron scale slice out of the single crystal. This small slice is then transferred to a sapphire substrate with ready-made gold contacts. To secure the slice to the substrate a small amount of epoxy is placed under the sample and cured for 1 h at 100 $^{\circ}$C.  Contacts are formed between the sample and the substrate's contacts by sputter-coating the sample and its surrounding region with 100 nm of gold. The FIB is then used once more to shape the TaAs slice into the final Hall bar configuration.

\noindent
\textbf{Photocurrent measurements.}
Output from a CO$_{2}$ laser operating at 10.6 $\mu$m is put through a mechanical chopper whose modulation frequency is computer controlled. The beam then goes through a quarter wave plate, whose angle is rotated via a motorized stage to control the light's polarization state. Finally, the light is focused onto the sample using a plano-convex ZnSe lens, giving a minimum spot size of $\sim$30\ $\mu$m. The sample itself is mounted on a 3-axis motorized translation stage providing fine tuning of the position of the sample within the laser spot. The generated photovoltages are measured by digital lock-in amplifiers on a National Instruments PXI DAQ system after amplification and frequency filtering via an SRS voltage pre-amplifier. The two point resistance of the contact pair is measured via current vs. voltage measurements, allowing conversion of the measured photovoltage to photocurrent. The photocurrents were fit using a function of the form:
\begin{equation}
J(\phi) = D + L_{S} \sin(4\phi) + L_{C} \cos(4\phi) + C_{S} \sin(2 \phi) + C_{C} \cos(2 \phi)
\end{equation}

\end{methods}

\begin{addendum}
\item Photocurrent experiments performed by G.O. were supported by the U.S. Department of Energy (DOE), Office of Science, Office of Basic Energy Sciences under Award No. DE-SC0018675. M.J.G. and K.S.B. acknowledge support from the National Science Foundation, Award No. DMR-1709987. L.K.D. was supported by a DAAD RISE fellowship. X.Y. and Y.R. acknowledge support from the National Science Foundation under Grant No. DMR-1151440. Work at UCLA was supported by the U.S. Department of Energy (DOE), Office of Science, Office of Basic Energy Sciences under Award No. DE-SC0011978. X.H. and P.J.W.M. were supported by the European Research Council (ERC) under the European Union's Horizon 2020 research and innovation programme (grant agreement No.~715730).
\item[Competing Interests] The authors declare that they have no competing financial interests.
\item[Author Contributions] K.S.B. and Y.R. conceived of the study. G.B.O. and L.K.D. performed the primary measurements, analyzed the data and provided the plots. M.J.G. performed the magneto-transport measurements. J.S., X.Y. and Y.R. performed the group theory analysis, LDA and tight binding calculations. G.B.O. and K.S.B. wrote the manuscript with input from all co-authors. B.S. and N.N. prepared the bulk crystals and P.J.W.M. and X.H. created the devices.
\item[Correspondence] Correspondence and requests for materials should be addressed to K.S. Burch.~(email: ks.burch@bc.edu).
\end{addendum}

\bibliography{References}

\end{document}